\documentclass[letterpaper,10pt]{article} 
\usepackage{opticameet3} %% use version 3 for proper copyright statement

\newcommand\authormark[1]{\textsuperscript{#1}}

\usepackage{amsmath,amssymb,listings,makecell,booktabs,multirow}
\usepackage{graphicx,subcaption}
\usepackage{tablefootnote}
\usepackage[colorlinks=true,bookmarks=false,citecolor=blue,urlcolor=blue]{hyperref} %pdflatex

\begin{document}

% \title{{RTRS}: A Practical Solution for Deployment of Generative Recommendations at Large-Scale}
% \title{SONIC: Serving Online Generative Recommendation at Large-Scale}
\title{FLAME: A Serving System Optimized for Large-Scale Generative Recommendation with Efficiency}

% \author{Author name(s)}
% \address{Author affiliation and full address}
% \email{e-mail address}

\author{Xianwen Guo, Bin Huang, Xiaomeng Wu, Guanlin Wu\authormark{*}, Fangjian Li, Shijia Wang, Qiang Xiao, Chuanjiang Luo, Yong Li}

\address{Netease Cloud Music, Hangzhou, China}

\email{\{guoxianwen, huangbin02, wuxiaomeng01, wuguanlin03, hzlifangjian, wangshijia1, hzxiaoqiang, luochuanjiang03, liyong02\}@corp.netease.com}

% \author{Author One,\authormark{1} Author Two,\authormark{2,*} and Author Three\authormark{2,3}}

% \address{\authormark{1}Publications Department, Optica, 2010 Massachusetts Avenue NW, Washington, DC 20036, USA\\
% \authormark{2}Peer Review, Publications Department, Optica, 2010 Massachusetts Avenue NW, Washington, DC 20036, USA\\
% \authormark{3}Currently with the Meetings Department, Optica, 2010 Massachusetts Avenue NW, Washington, DC 20036, USA}

% \email{\authormark{*}cstech@optica.org} %% email address is required

%% Do not add a copyright statement. Optica will add it.

\begin{abstract}
Generative recommendation (GR) models possess greater scaling power compared to traditional deep learning recommendation models (DLRMs), yet they also impose a tremendous increase in computational burden.
{Measured in FLOPs, a typical GR model's workload sits in $10^9 \sim 10^{11}$ range, roughly four orders of magnitude higher than traditional DLRMs.
Delivering accurate results in a few tens of milliseconds while processing billions of such requests per day puts extreme demands on the performance of the online serving system.}
Therefore, for industry practitioners, the alluring gains of GR models are tempered by the formidable challenge of online deployment at scale in production services.
% For major industry players, the attractive gains of GR models come with a steep challenge: serving them at scale under massive online-request loads.
{In this work, we introduce a comprehensive solution of online serving system tailored \textbf{F}or \textbf{L}arge-scale Gener\textbf{a}tive Reco\textbf{m}mendation with \textbf{E}fficiency (FLAME).}
% In this work, we present a comprehensive inference system tailored for large-scale online deployment of GR models, leveraging CPU-GPU heterogeneous hardware environments. 
Specifically, we leveraging CPU-GPU heterogeneous hardware to decouple feature pre-processing and model computation.
{We encapsulated several memory optimization features as the Proximal Data Accelerator (PDA) module to make full use of limited bandwidth and storage resources, which achieves a 1.9x throughput gain and a 1.7x latency reduction.}
We implement the Fused Kernel Engine (FKE) module based on the functionality and interface of NVIDIA TensorRT to boost model computation, delivering a speedup ratio of 4.6x-6.1x, throughput gain ratio of 4.7x-6.3x one step further.
In addition, we design the Dynamic Stream Orchestrator (DSO) module to coordinate concurrent requests, enhancing the system throughput performance {with 1.3x improvement in throughput and 2.3x speed-up under non-uniform distribution of upstream candidates.}
Comprehensive evaluations demonstrate that our FLAME effectively supports large-scale online deployment of GR models and achieves remarkable improvements in system performance.

\end{abstract}

\section{Introduction}
With the continuous development and expansion of the Internet of Things (IoT), the scale of big data is still growing at an astonishing rate\cite{bansal2020survey}.
Identifying the suitable items from massive data instantaneously to match precisely with user preference is a challenge for recommendation systems. 
It's also the key for major application manufacturers to enhance user experience and platform revenue. 
To cope with the challenges of multi-scenario and multi-task recommendation tasks, traditional deep learning recommendation models (DLRMs) rely heavily on cumbersome feature engineering and manually customized neural networks \cite{cheng2016wide,guo2017deepfm,wang2017deep,zhou2018deep,zhou2019deep,tang2020progressive}.
\par
Recently, large language models (LLMs) have demonstrated near-human intelligence in some specific scenarios, showing excellent performance in tasks such as logical reasoning in mathematics and coding \cite{openai2025gpt4.5,comanici2025gemini,guo2025deepseek,yang2025qwen3,team2025kimi}. 
The continuous emergence of new scaling laws in the research of LLMs \cite{kaplan2020scaling,hoffmann2022training,wu2024inference,shao2024scaling,busbridge2025distillation} provide practical guides to improve model performance with affordable resource budgets .
Scaling laws in the recommendation field have gradually emerged from obscurity to concreteness in recent studies \cite{zhang2024wukong, zhai2024actions, wang2024scaling, lv2024marm}.
Notably, the paradigm of generative recommendation (GR) \cite{zhai2024actions} demonstrates greater scaling power compared to traditional DLRMs through its streamlined feature engineering and unified sequence model.
Thanks to the network architecture dominated by attention mechanism with large amount of parameters, GR models have higher potential to capture the complex dynamics in user interaction sequences.
Recent studies have successfully implemented GR models in various application fields \cite{xu2025efficient, huang2025towards, han2025mtgr, deng2025onerec, zhou2025onerec}, providing solid cases for the conception of the next-generation recommendation systems.

As one of the largest music streaming platforms and user communities in China, Netease Cloud Music recently launched the Climber GR model and achieved remarkable benefits in both offline experiments and online A/B tests \cite{xu2025efficient}. 
{In line with typical production-environment settings, operating conditions and system requirements for an online serving system are demonstrated in Table \ref{table:intro}.}
From this, it's clear that deploying GR models on large-scale real-time online inference services faces tremendous challenges:

\begin{itemize}
\item \textbf{Low latency.} 
Online recommendation services need to make accurate responses to users within an extremely short period of time (i.e., dozens of milliseconds). 
Taking the ranking stage as an example, the time-consuming per request can be broken down into the processes of feature querying, feature assembly, model computation, and response packaging. 
Among these, model computation will pose a huge burden for GR models. 
A large number of attention modules are used to process long user interaction sequences, and the leading term of computational complexity is proportional to the square of the sequence length. 
Model computation measured by FLOPs for a single request reaches $10^9 \sim 10^{11}$, which is roughly four orders of magnitude higher than that of traditional DLRMs. 
Therefore, how to complete user request instantaneously poses higher challenges for the optimization of computing kernels and the utilization of computing resources.

\item \textbf{High throughput.} 
The current online recommendation service needs to handle $10^{10} \sim 10^{12}$ requests from tens of millions of users every day. 
Such large-scale requests will compete for priority computing resources (data storage and data transfer bandwidth, etc.). 
If not handled properly, it will lead to system performance degradation and even affect user experience in serious cases.

\item \textbf{Efficient utilization of heterogeneous resources.} 
To meet the high computational complexity of GR models, we have introduced clusters containing GPU hardware modules to accelerate model computation. 
Thus, the entire online serving system is built on a heterogeneous CPU-GPU  hardware environment. 
Due to the inherent difference between CPU and GPU in terms of storage structure, data transfer, and instruction execution, special considerations and designs are required to make the whole system work at its best.
\end{itemize}

\begin{table}[htbp]
 \centering 
    \caption{Typical operating conditions and system requirements for an online serving system.}
    \begin{tabular}{cccc}
    \hline
    Model FLOPs & Hardware Peak Performance & Overall Latency & Requests per Day \\
    \hline
    \makecell[l]{$10^6\sim 10^7$ (DLRMs)\\$10^9 \sim 10 ^{11}$ (GR)} &
    \makecell[l]{$403.2\times 10^9$ (CPU) \tablefootnote{The CPU example here is Intel(R) Xeon(R) Gold 5220R 2.20 GHz. Detailed specifications can be found at https://cdrdv2-public.intel.com/840270/APP-for-Intel-Xeon-Processors.pdf} \\ $19.5\times 10^{12}$ (GPU) \tablefootnote{The GPU example here is NVIDIA A100 Tensor Core GPU. Detailed specifications can be found at https://www.nvidia.com/content/dam/en-zz/Solutions/Data-Center/a100/pdf/nvidia-a100-datasheet-us-nvidia-1758950-r4-web.pdf}}
    & $<50$ ms & $10^{10} \sim 10^{12}$\\
    \hline
    \end{tabular}
\label{table:intro}
\end{table}

% \begin{figure}[htbp]
%   \centering
%   \includegraphics[scale=0.5]{fig/decouple.pdf}
% \caption{A typical workflow of online serving system under CPU-GPU heterogeneous hardware environment.}
% \label{fig:decouple}
% \end{figure}

\begin{figure}[htbp]
  \centering
  \includegraphics[scale=0.45]{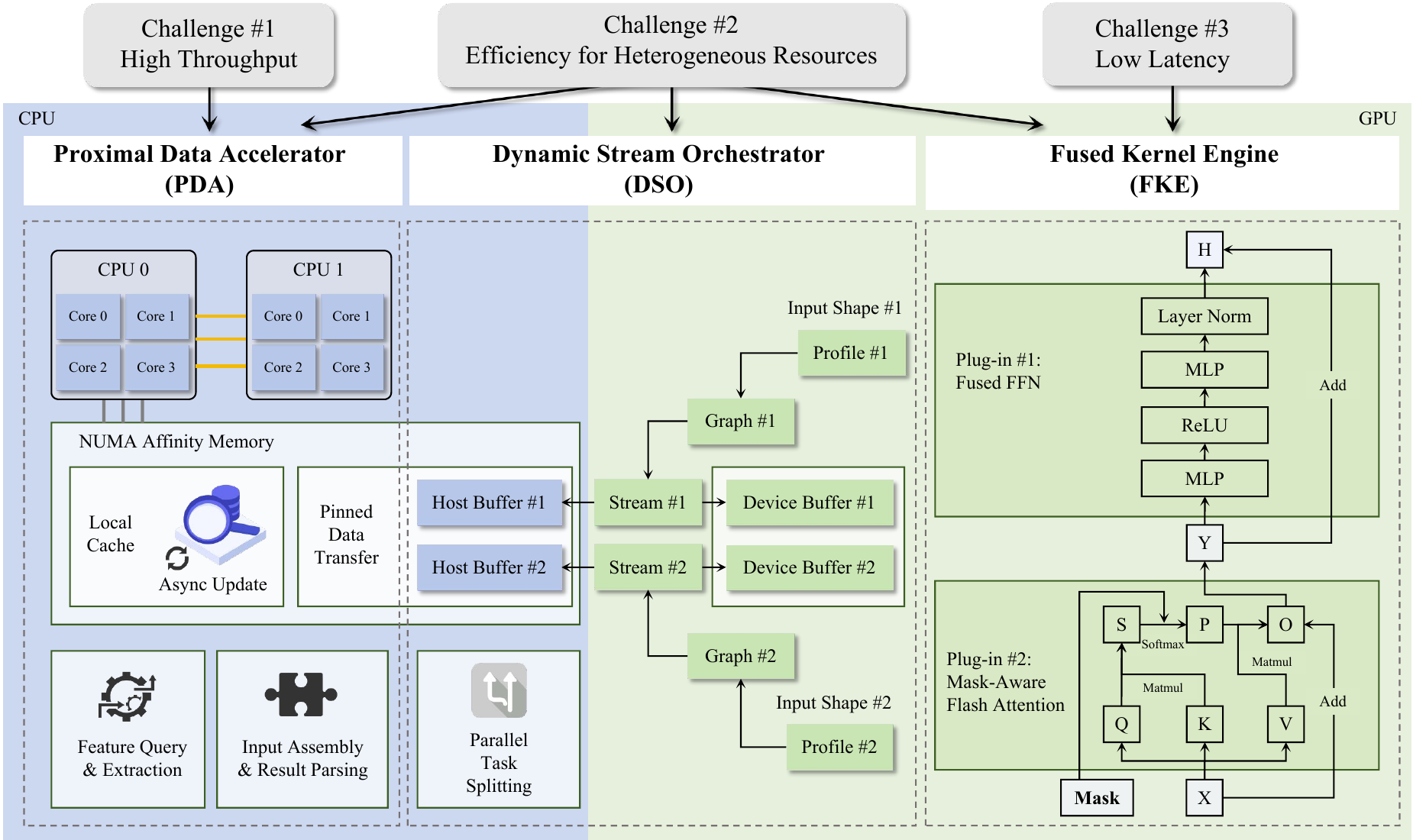}
\caption{System Overview. FLAME adopts a decoupled architecture to separate feature processing with model computation on heterogeneous hardware.}
\label{fig:overview2}
\end{figure}

To address these challenges, we introduce a comprehensive solution of online serving system optimized \textbf{F}or \textbf{L}arge-scale Gener\textbf{a}tive Reco\textbf{m}mendation with \textbf{E}fficiency (FLAME). 
The overview of the whole system is demonstrated in Figure \ref{fig:overview2}.
The whole system adopts a decoupled architecture to take advantage of the distinct inherent characteristics of CPUs and GPUs. 

We construct the feature-processing engine on CPU to handle all the operations before model computation, such as feature querying from remote servers, customized feature extraction operations (e.g. type conversion, embedding look-up), and input assembly. 
Specifically, we encapsulated several memory optimization features as the proximal data accelerator (PDA) to make full use of limited bandwidth and storage resources. 
We introduce asynchronous feature querying and multi-level caching mechanisms to reduce the network transmission overhead between remote services and the local system. We apply the core binding techniques for modern CPU with NUMA (Non-Uniform Memory Access) architecture to improve the efficiency of CPU memory access. 
We utilize pinned memory to enhance the efficiency of data transfer from host to device.

Based on the high performance parallel computing power of GPU, we build the fused-kernel engine (FKE) based on TensorRT library\cite{TensorRT}. 
We implement forward computation of the whole model through TensorRT API for optimized performance instead of ONNX format model conversion by default. 
In addition, we replace the attention module and FFN (Feed-Forward Network) module by customized plug-ins to achieve kernel fusion.
To cope with the demand of ``single user, multiple items(SUMI)" paradigm per request during the inference phase, 
% we adopt the two-stage forward computaion with key/value cache vectors as intermediate variables like M-FALCON\cite{zhai2024actions}. 
we adapt the prestigious Flash-Attention\cite{dao2022flashattention, dao2023flashattention} in a mask-aware style as a TensorRT plug-in to replace the default attention operator. 
For matrix multiplications within the above kernels, we implemented pipeline
processing between data loading and general matrix multiplication (GEMM) via CUTLASS\cite{Thakkar_CUTLASS_2023}.

Further, we design the dynamic stream orchestrator (DSO) module to coordinate concurrent requests, enhancing performance to the next level.
DSO implements dynamic batch routing by exporting multiple TensorRT optimized profiles with specified batch shapes during engine construction. 
Each profile is equipped with preallocated input/output buffers (based on preset batch dimensions) and CUDA Graph-captured computation processes (for efficient kernel execution). 
These components form \verb|executor| managed via an index queue. 
Upon upstream requests, tasks are dynamically split by batch size (in descending order), assigned to available executors, and returned to the queue post-computation, thereby enhancing system throughput.

In summary, this paper makes the following contributions:
\begin{enumerate}
\item We provide a comprehensive practical solution of large-scale online inference optimized for generative recommendation service under heterogeneous hardware environments. 
\item We innovatively implement the PDA, FKE, and DSO design modules. By organically integrating the advantageous characteristics of the hardware, these modules form a powerful link in accelerating the performance of the entire inference service, providing a solid reference case for similar industry applications.
\item We conducted a series of performance tests and analyzed the overall performance impact of each module. These supported the deployment of GR model in achieving excellent performance in real online application scenarios.
\end{enumerate}

\section{Background}
\subsection{A Brief Review on GR model in Netease Cloud Music}
\begin{figure}[htbp]
  \centering
  \includegraphics[width=9cm]{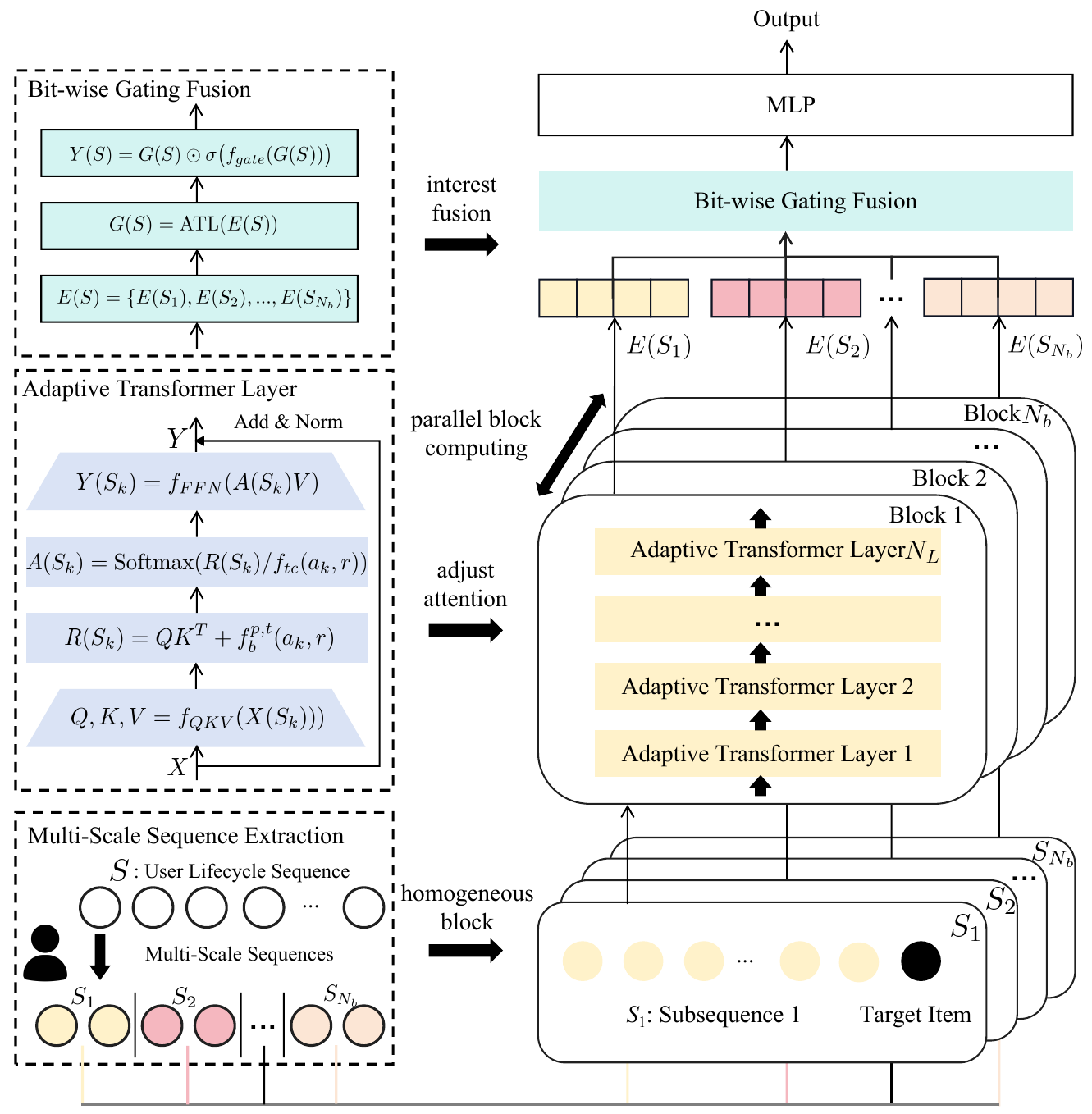}
\caption{Network architecture of the Climber GR model\cite{xu2025efficient}.}
\label{fig:climber}
\end{figure}

Before introducing the inference system, let's first briefly review our GR model, Climber\cite{xu2025efficient}. Overall, the model adopts a Transformer-dominated\cite{vaswani2017attention} architecture to make predictions based on the sequence of items related to user's historical behaviors. 
To alleviate the computational overhead of self-attention brought by excessive length, user sequences are reorganized into several sub-sequences which are processed by different independent Transformer blocks. 
The number of blocks can be denoted as $N_b$. 
In this design, the computational complexity related to attention is reduced from $O(n^2 d)$ to $O(n^2 d / N_b)$,  where $n$ and $d$ denote the total length of original user behavior sequence and the hidden dimension. 
Within each block, an adaptive temperature coefficient is introduced before softmax activation to adapt to different segmentation strategies and the multi-scenario nature of recommendations. 
The target items will be concatenated as the last element with user's behavior sequence of each block, and the output results of each block will then be fused through the bit-wise gating fusion module. 
Finally, the probability scores across multiple tasks can be obtained through the expert module of the top-level multi-layer perceptron (MLP). The overall architecture of model is demonstrated in Figure \ref{fig:climber}.
The above model achieves performance scalability with increasing computational FLOPs in both offline training metrics and online business metrics through low-traffic A/B online experiments. This also motivated a series of engineering optimization efforts in the serving system, supporting the GR model from low-traffic validation to large-scale online deployment.

\subsection{Hardware Environment}
\begin{figure}[htbp]
  \centering
  \includegraphics[width=15cm]{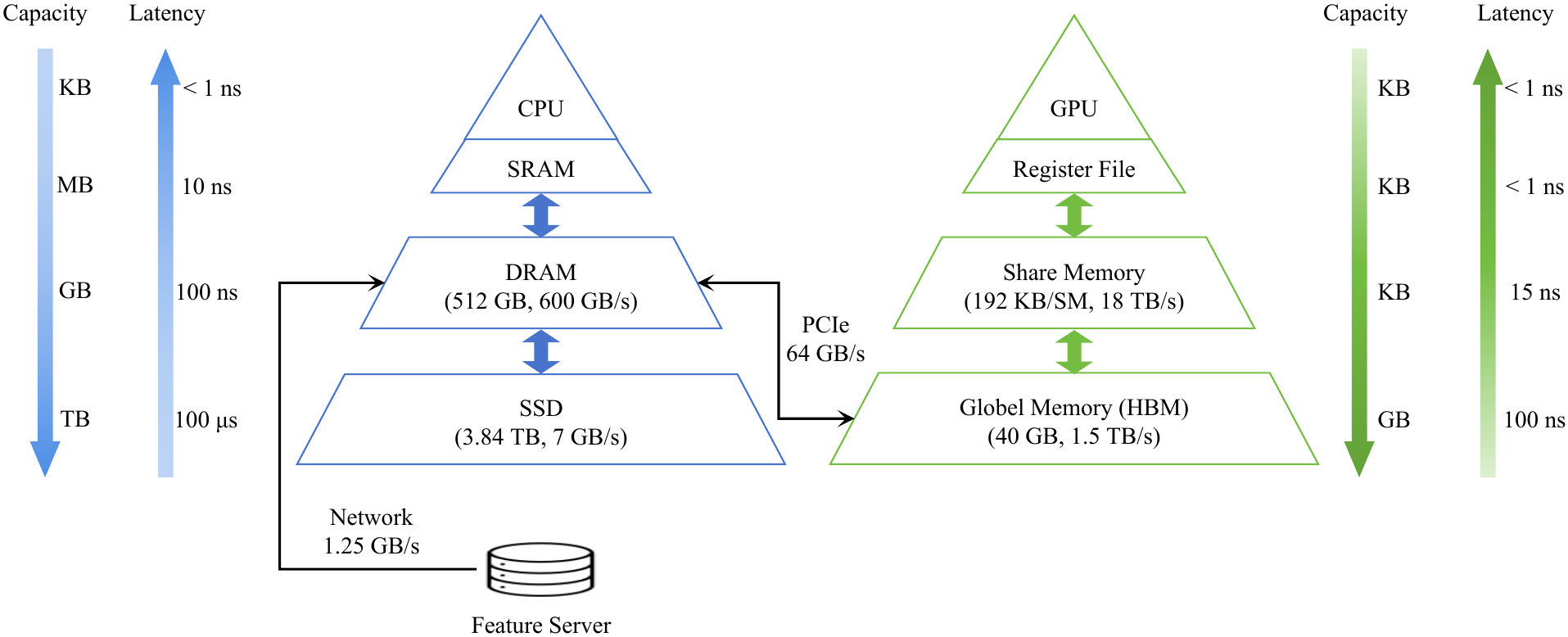}
\caption{Data flow and typical bandwidth between heterogeneous hardware components. The specification of the GPU part referred to NVIDIA A100 \cite{Peter2021Dissecting}.}
\label{fig:hardware}
\end{figure}

CPU has been a popular choice for DLRM inference due to its relatively economical cost and ease of availability\cite{Gupta2020DeepRecSys,Gupta2020Arch,Ke2022Hercules,Liu2021Understanding,liu2021jizhi,Jain2023Optimizing}. 
However, the dramatic increase in computational load of the GR model might exceed the upper limit of hardware computing performance that pure CPUs can provide for large-scale serving systems, necessitating a shift to CPU-GPU heterogeneous hardware architectures. 
In this setup, the core function of the inference system is to achieve efficient data flow to handle high-concurrency requests. 
Therefore, it is vital for systematic optimization to understand the bandwidth performance of each hardware component involved in the entire data flow process.
The entry point of the serving system is the receipt of a network request uniquely identified by the user ID, with common network bandwidth typically being 1.25 GB/s. 
After obtaining the user identifier, the candidate item identifier, and context information from the request, a remote feature query service is usually invoked to prepare the input for the model computation. 
Results from feature query are stored in CPU memory for further operation like array concatenation, type casting, etc.  
After the model input is ready, these data will be copied to the GPU's global memory via the PCIe bus for subsequent parallel computing. 
We take the NVIDIA A100 Tensor Core GPU as an example.
Inside the GPU, there is also a multilevel hierarchical storage architecture, ranging from the global memory to the shared memory of each SM (Streaming Multiprocessor), and finally to the register file closest to the computing cores. 
A common parallel computing strategy is to break down large-scale computing tasks into independent small-scale tasks. 
By progressively distributing the data to the on-chip register files closest to the computing units, high-performance computing can be achieved using CUDA cores or tensor cores. 
After completing the computing tasks and data synchronization, the results are copied from the GPU back to the CPU, where the response body is packaged to complete the entire data flow. 
The entire data flow and bandwidth corresponding to the hardware component are shown in Figure \ref{fig:hardware}. 

\subsection{System Overview}
Finally, in conjunction with the hardware setup, let's take a look at the overall architecture of the FLAME serving system. 
To fully exploit the advantage of hardware modules, we adopted a decoupling architecture to separate feature processing from model computation. 
The part of CPU handles all feature processing operations for input assembly, embedding look-up and result parsing. The GPU part is only responsible for the remaining model computation.

\begin{figure}[htbp]
  \centering
  \includegraphics[scale=0.45]{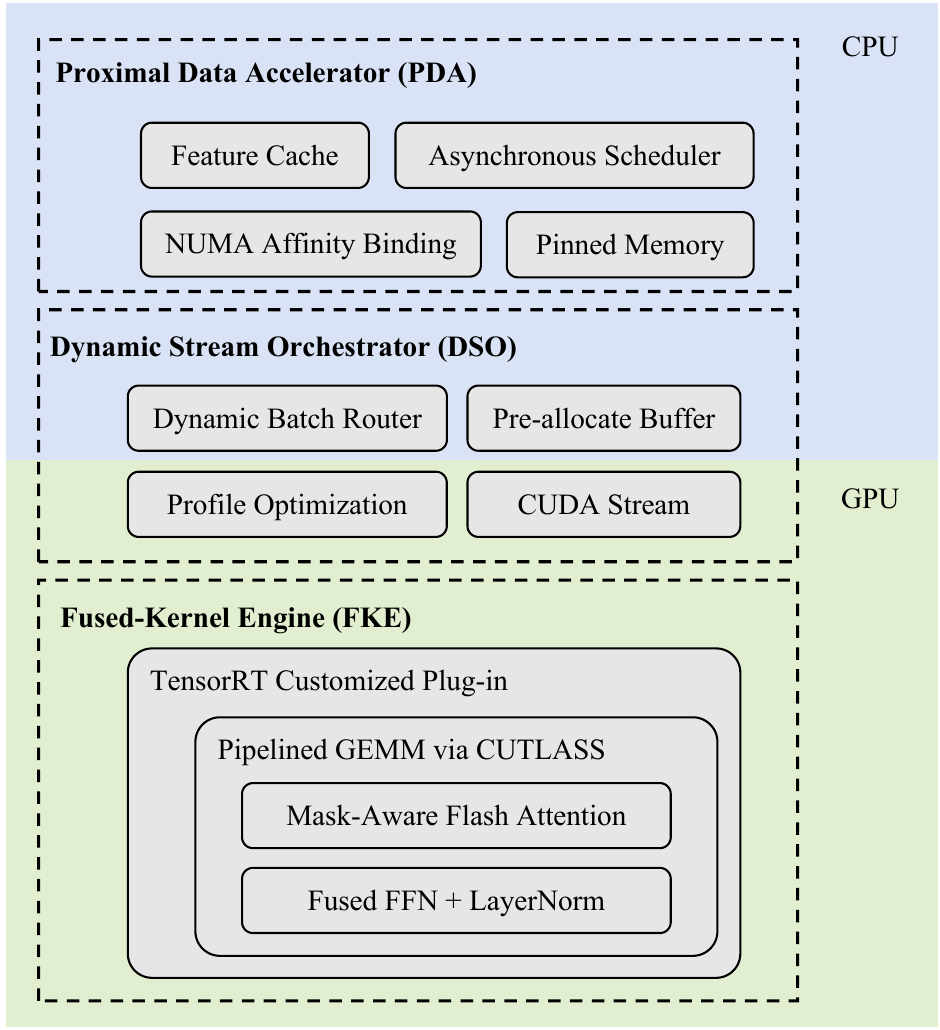}
\caption{System Overview}
\label{fig:overview}
\end{figure}

Firstly, we designed the proximal data accelerator module (PDA) to improve the efficiency and throughput of data pre-processing. 
The PDA module contains three core mechanisms: asynchronous feature query with cache, NUMA affinity core binding, and pinned data transfer. 
Secondly, to fully utilize the GPU computing power for our GR model, we implemented the fused kernel engine (FKE) based on the TensorRT interface. 
We rebuild the forward network through TensorRT API as the inference engine. 
With the aid of the TensorRT plug-in mechanism, we adapt the prestigious Flash-Attention\cite{dao2022flashattention, dao2023flashattention} in a mask-aware style to replace the default attention module to form a fused kernel. 
We also fused layer normalization and linear projection into another plug-in.
% During the construction of the inference engine, CUDA graph optimization is carried out. 
Furthermore, we design the dynamic stream orchestrator (DSO) module to coordinate the overall utilization of resources. 
The CUDA stream mechanism is employed to interleave the processing efficiency of multiple requests and to overlap the data transfer with model computation.
To address the uncertainty in the number of candidate items requested by upstream services, we utilized the explicit shape mode during the building stage of the TensorRT inference engine. 
We exported profiles containing multiple shapes and preallocated memory buffers. 
With batch routing algorithm, requests under various number of candidates can be delivered to proper executors.
In summary, the key technical elements of the entire inference system are shown in Figure \ref{fig:overview}.

\section{Methodology}
\subsection{Proximal Data Accelerator}
\begin{figure}[htbp]
  \centering
  \includegraphics[scale=0.5]{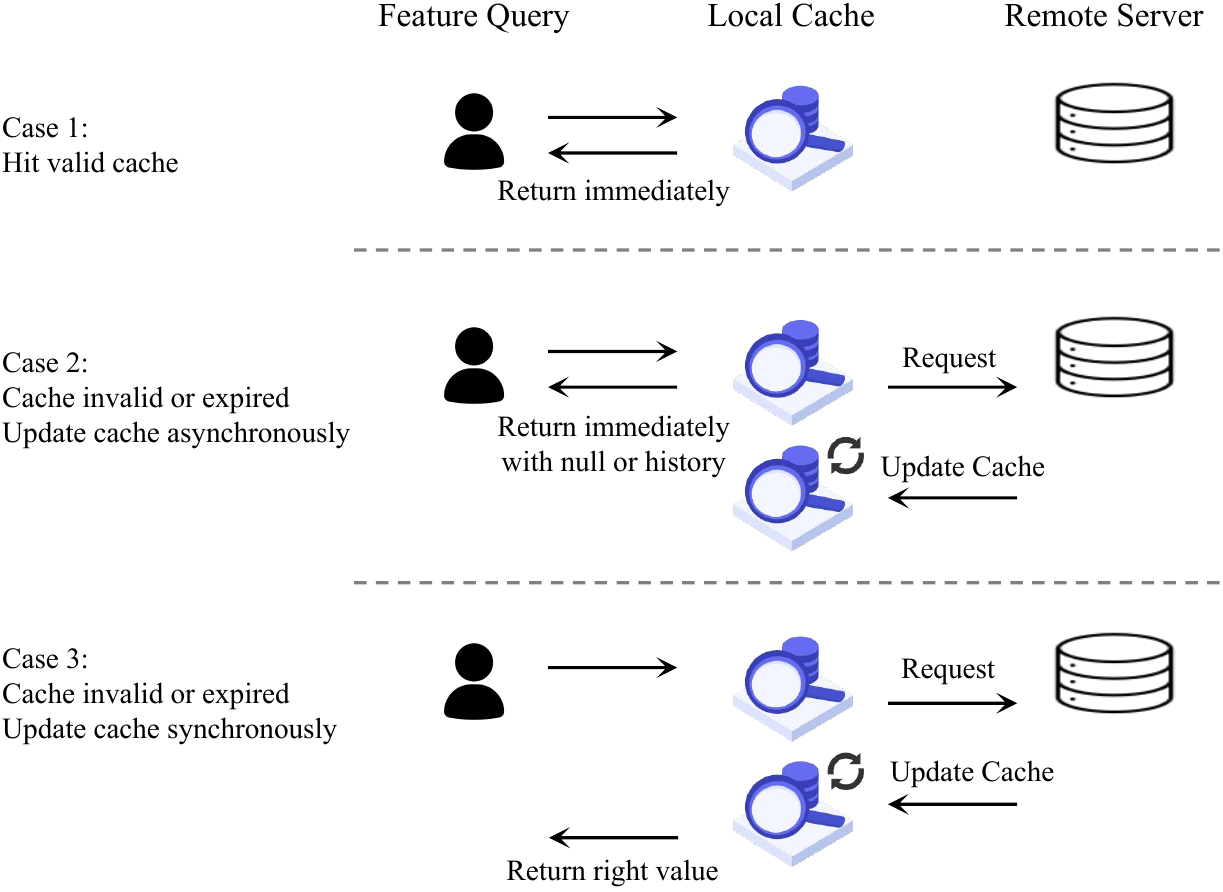}
\caption{Feature query with cache}
\label{fig:query}
\end{figure}

As we mentioned above, the bandwidth for data transfer in local memory is hundreds times higher than that of network transmission bandwidth. 
To avoid network bandwidth becoming a bottleneck for overall performance, the first key design of the PDA module is a feature query engine with caching mechanism. 
{Considering the characteristics of the music platform recommendation business, introducing a caching mechanism for feature queries on the core hot items side offers greater benefits compared to caching feature on the user side.}
When a query request for an item is triggered, the object cache within the process is first queried. 
If an unexpired cache hit occurs, the result is returned directly. 
If a cache hit occurs but the cache is expired, the expired result is returned directly, while an asynchronous query task is initiated. 
This task runs in the background and does not block waiting for the query result to return. 
Once the asynchronous query task is completed, the local cache is updated. 
If no cache hits occur, an empty result is returned, and the same asynchronous query task is initiated.
The aforementioned asynchronous caching can provide high query performance, but in cases where the query does not hit the cache, it can lead to missing features, which in turn affects the accuracy of the recommendation results. 
Therefore, we have also implemented a synchronous query mechanism with caching. 
Compared to asynchronous querying, when a cache miss occurs or the cached result is expired, a synchronous query request will be initiated, waiting for the result to return and updating the local cache. 
Figure \ref{fig:query} demonstrates feature query with cache in asynchronous and synchronous settings.
This synchronous query mechanism ensures the accuracy of the results. 
The cache object adopts an LRU (Least Recently Used) design, and in terms of storage structure, it can be divided into multiple buckets to reduce write lock collisions and further enhance query performance.

\begin{figure}[htbp]
  \centering
  \includegraphics[scale=0.45]{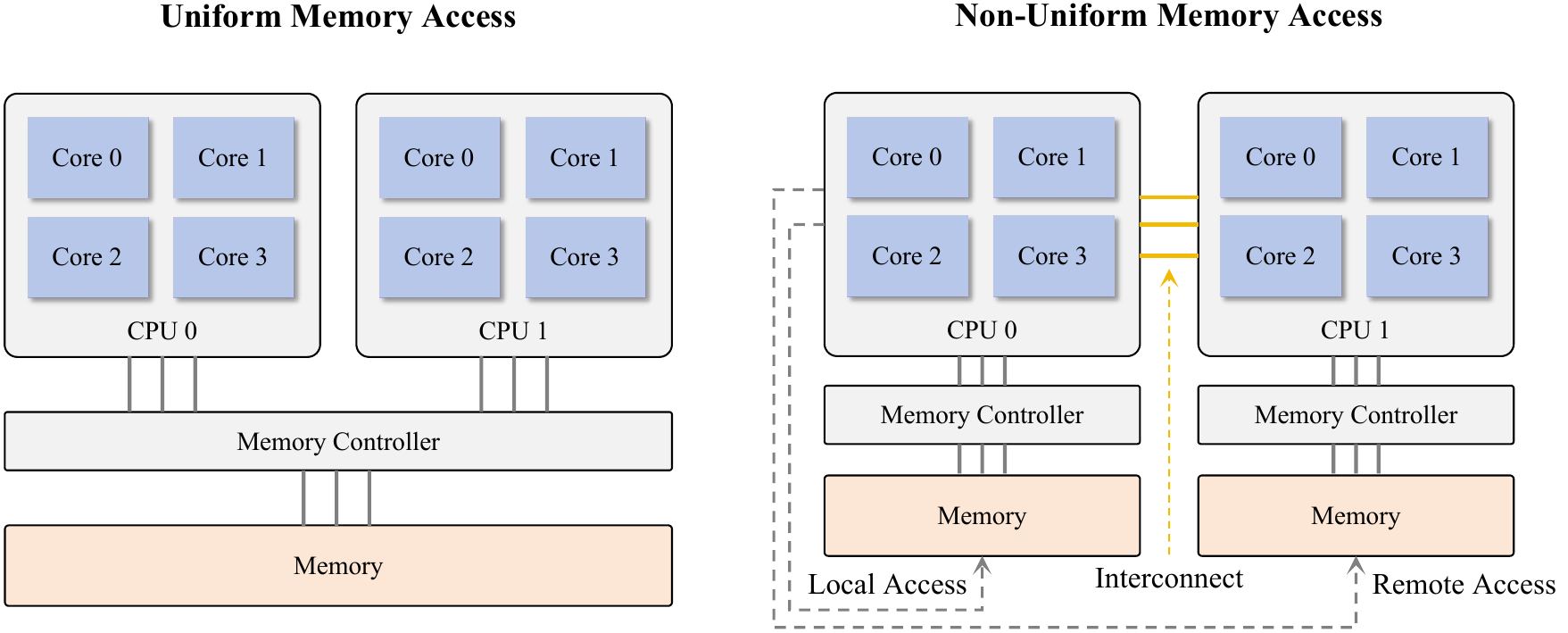}
\caption{Demonstration of UMA and NUMA architecture}
\label{fig:numa}
\end{figure}

The second key setting of the PDA module is NUMA core binding. 
The computing power of single-core processors can no longer meet the growing computational demands, leading processor manufacturers to develop multi-core processors. 
The architecture of single-machine servers has evolved from Uniform Memory Access (UMA) to Non-Uniform Memory Access (NUMA).
In Uniform Memory Access (UMA) architectures, all processors share a single, centralized memory pool interconnected via a common bus or switch fabric. 
This design ensures that memory access latency remains consistent across all processing units, as every request traverses the same shared pathway to the unified memory resource. 
However, this uniformity comes at the cost of scalability: the shared interconnect becomes a critical bottleneck as the number of processors increases.
In contrast, Non-Uniform Memory Access (NUMA) architectures address scalability limitations by partitioning memory into discrete ``nodes," each collocated with a subset of processors and connected via high-speed inter-node links\cite{Frank2016NUMA}. 
This topology introduces inherent latency asymmetry: processors access "local" memory within their node with significantly lower latency than "remote" memory in other nodes, which must traverse inter-node interconnects. 
By distributing memory and processing resources, NUMA mitigates the bottlenecks of shared buses, enabling support for large-scale systems with dozens to hundreds of cores -- common in high-performance computing (HPC) clusters and cloud infrastructure. 
However, this advantage necessitates NUMA-aware optimizations in both operating systems (e.g. affinity-based scheduling) and applications (e.g., localized memory allocation) to minimize performance penalties associated with frequent remote memory accesses. 
We can use the \verb|numactl| tool to bind processes to specific nodes, or directly bind threads using the C++ \verb|pthread_attr_setaffinity_np| interface. By effectively utilizing NUMA core binding technology, we can significantly enhance the performance of servers under high-load scenarios. Figure \ref{fig:numa} shows the difference in memory access between UMA and NUMA architecture. 

\begin{figure}[htbp]
  \centering
  \includegraphics[scale=0.45]{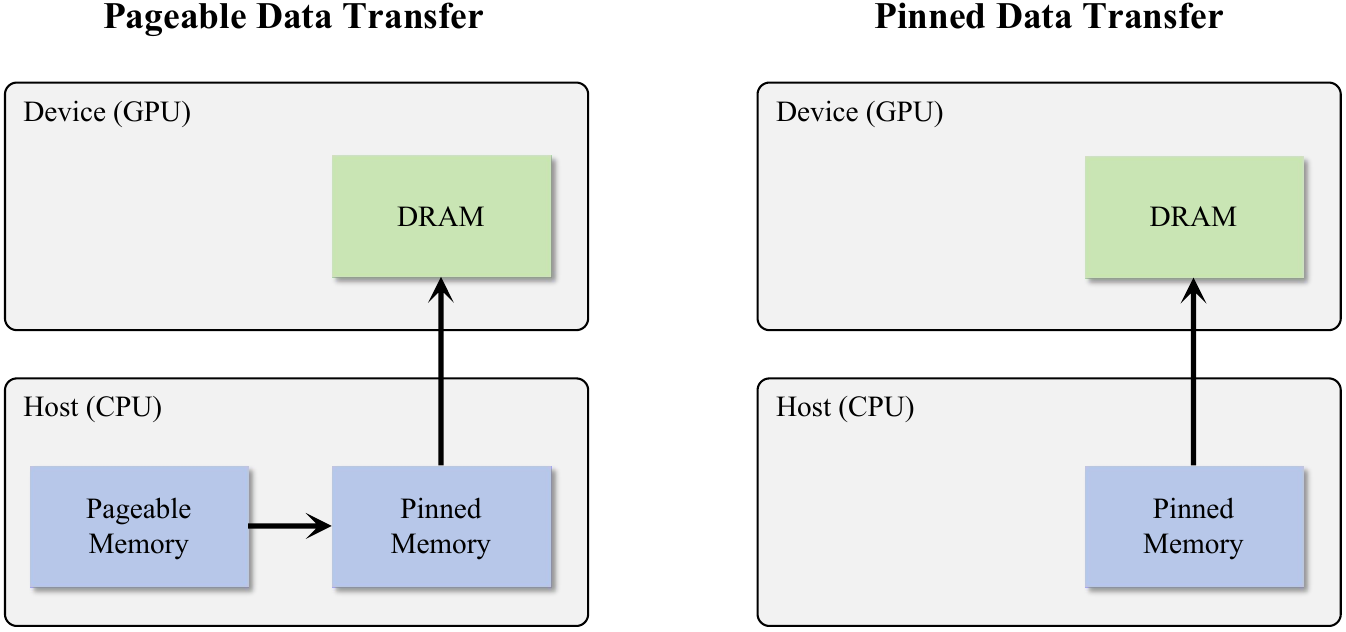}
\caption{Demonstration of pageable data transfer and pinned data transfer\cite{Mark2012CUDA}.}
\label{fig:pinned}
\end{figure}

The last key point of the PDA module is the use of pinned memory. 
When performing CUDA programming under a heterogeneous hardware architecture, memory space is first allocated on the host side (CPU) to store model input variables, and then the data are transferred to the device side (GPU) via the PCIe bus, as shown in Figure \ref{fig:hardware}. 
However, host data allocations are pageable by default and cannot be directly accessed by the GPU. 
So when a data transfer from pageable host memory to device memory is invoked, the CUDA driver must first allocate a temporary page-locked, or “pinned”, host array, copy the host data to the pinned array, and then transfer the data from the pinned array to device memory, as illustrated in Figure \ref{fig:pinned}. 
To avoid additional memory usage and data copying, we use the \verb|cudaMallocHost| or \verb|cudaHostAlloc| function offered by CUDA C++ interface to directly allocate pinned host memory\cite{Mark2012CUDA}.  
Further, model input variables can be packaged as a whole to batch many small transfers together into a single transfer. 

\subsection{Fused-Kernel Engine}
We developed the Fused Kernel Engine (FKE) core computing module based on the functionality and interface of NVIDIA TensorRT, which is an SDK for high-performance deep learning inference on NVIDIA GPUs\cite{TensorRT}. 
The basic workflow typically involves converting models in the ONNX format to a singular TensorRT inference \textit{Engine} file, where the ONNX format enables computational compatibility independent of the training framework. 
This offers a convenient and largely automated route for translating trained deep-learning models into TensorRT engines. 
However, this convenience comes at the cost of certain structural constraints. 
Specifically, the ONNX–TensorRT bridge demands that every operation present in the source graph be either natively supported by TensorRT or explicitly re-implemented through custom plug-ins; any deviation aborts the conversion. 
Furthermore, the sequence of graph-level optimizations executed by the ONNX parser remains opaque to the practitioner, occasionally yielding fusions whose efficiency falls short of what could be achieved through deliberate design. 
In architectures of exceptional depth or unconventional topology the like GR models, the exported ONNX representation may become gratuitously verbose, impeding TensorRT’s ability to compress memory footprints or to expose maximal parallelism.

\begin{figure}[htbp]
  \centering
  \includegraphics[scale=0.45]{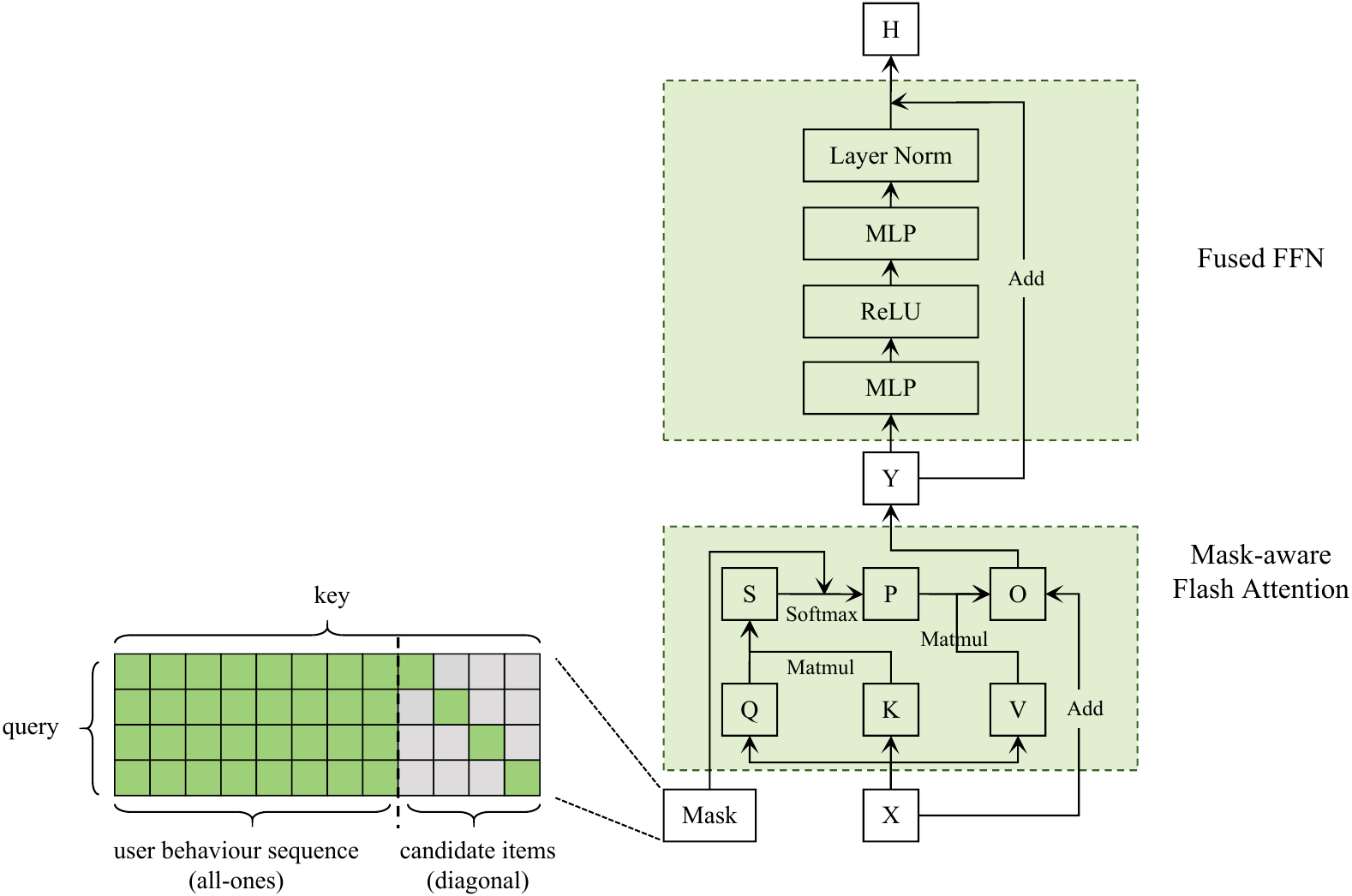}
\caption{Kernel fusion for attention module and FFN module. We implemented mask-aware flash attention kernel for parallel prediction of candidate items.}
\label{fig:fusion}
\end{figure}

\begin{figure}[htbp]
  \centering
  \includegraphics[scale=0.65]{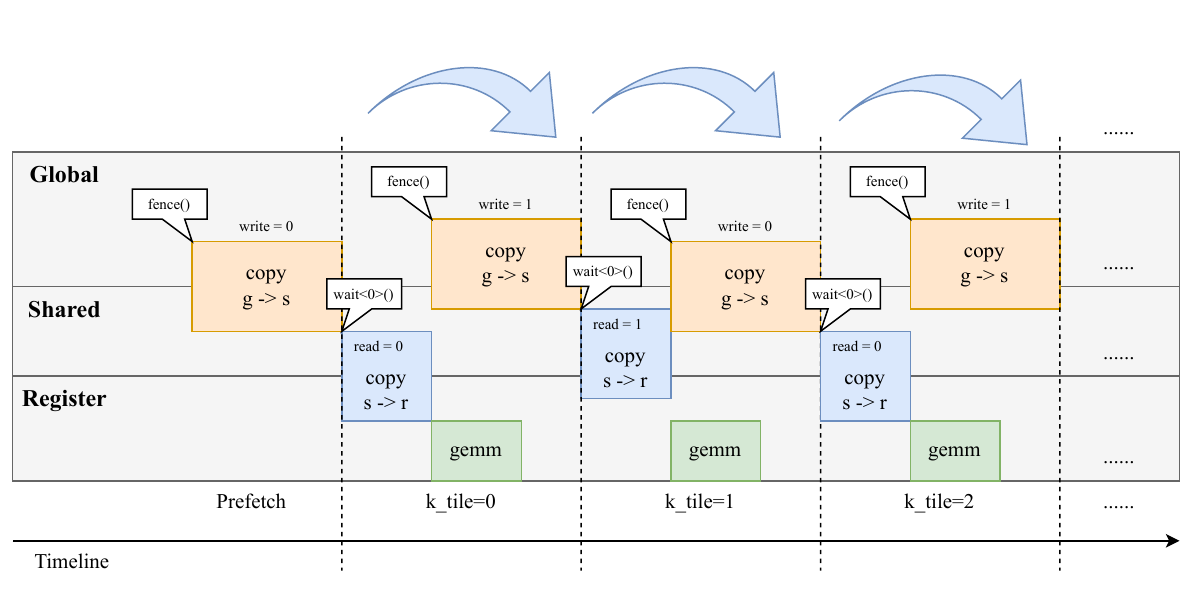}
\caption{Pipeline matrix multiplication overlapping data transfer across GPU's hierarchical memory architecture and GEMM operations.}
\label{fig:gemm}
\end{figure}

Consequently, the TensorRT network definition API retains its relevance as a complementary avenue. 
By constructing the network layer-by-layer, the developer gains explicit jurisdiction over operator selection, tensor layout, precision calibration, and fusion strategy. 
Thus, to achieve the best performance, we constructed the GR model based on the TensorRT API layer by layer and implemented the fused operators for attention computation and the subsequent fusion of the feed-forward network (FFN) and layer normalization modules in the form of plug-ins, as shown in Figure \ref{fig:fusion}.
Specifically, we refer to the implementation of Flash-Attention to fuse the attention module\cite{dao2022flashattention, dao2023flashattention}.  
On this basis, we added processing for customized mask matrices to achieve parallel computation of the prediction scores for candidate items like HSTU \cite{zhai2024actions}. 
We refer to the implementation of customized masks in the HSTU kernel from NVIDIA's RecSys Example\cite{nvidia2024rec}. 
By using the interfaces provided by CUTLASS\cite{Thakkar_CUTLASS_2023}, we located the two-dimensional coordinates of each thread's corresponding values in the result matrix during the $QK$ matrix multiplication. 
By determining whether these coordinates fall within the mask boundary, we can decide whether local elements need to be masked. 
During the matrix multiplication process, we utilized the asynchronous copy instruction \verb|cp_async| of the NVIDIA Ampere architecture to achieve pipeline processing between data loading and general matrix multiplication (GEMM) as shown in Figure \ref{fig:gemm}. 

\subsection{Dynamic Stream Orchestrator}
\begin{figure}[htbp]
  \centering
  \includegraphics[scale=0.65]{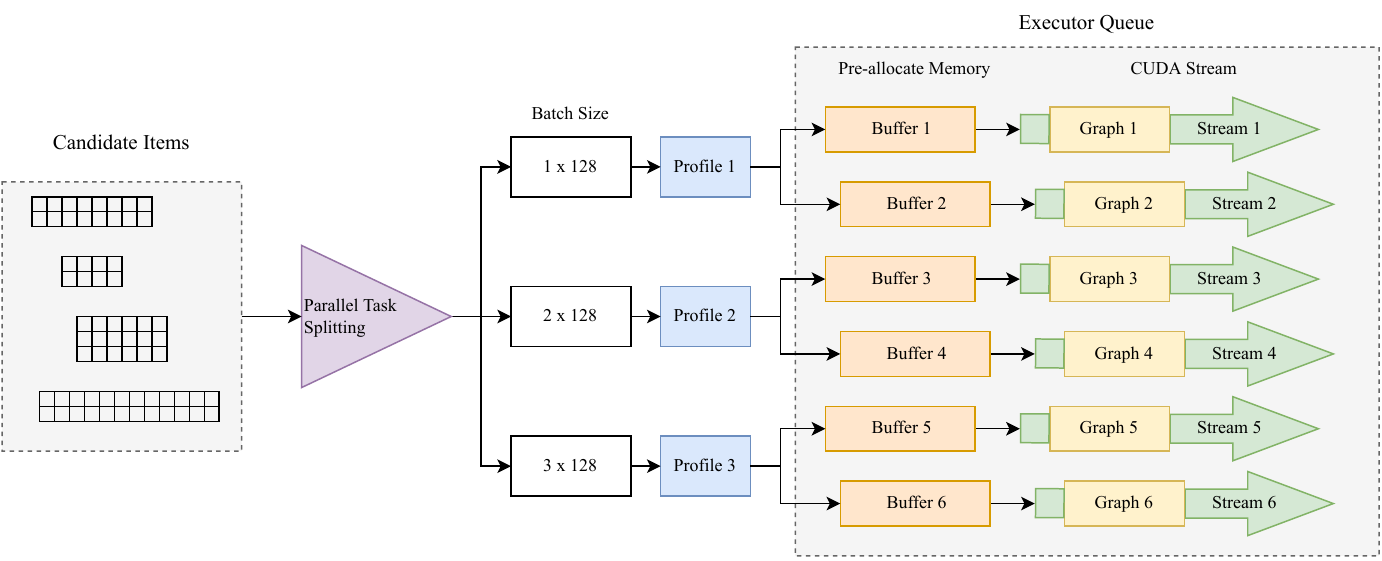}
\caption{The schematic diagram of the DSO module.}
\label{fig:dso}
\end{figure}

We proposed the Dynamic Stream Orchestrator (DSO) module, which leverages the TensorRT profile optimization mechanism and the CUDA stream scheduling mechanism to further enhance the system's throughput performance in cases where the distribution of candidate items from upstream services is nonuniform. 
A common practice is to set certain dimensions to -1 for cases where the dimensions of input variables cannot be predetermined. 
This allows for dimension inference at runtime while also triggering temporary dynamic memory allocation. 
However, this approach leads to frequent memory allocation and de-allocation, significantly increasing the overall processing time. 
Moreover, the varying sizes of each allocation inevitably lead to memory fragmentation. 
The uncertainty in computational dimensions also prevents the use of CUDA Graph to package and optimize kernels and other operations.
For the reasons mentioned above, we propose the DSO to implement the functionality of dynamic batch routing. 
Firstly, when building the TensorRT engine, multiple optimized profiles with specified batch shapes can be exported. 
Each profile can be equipped with multiple CUDA streams, and the corresponding storage buffers for holding input and output variables can be pre-allocated based on the pre-set batch dimensions. 
Another advantage of fixing batch dimension is that the model computing process can be captured into a CUDA graph during initialization. 
Subsequent requests can then directly invoke the execution of the CUDA graph to benefit from the efficient merging of kernel launches and other operations. 
Each corresponding buffer, stream, and graph form a set of core components, which we define as an \verb|executor|, and we maintain an \verb|executor| index queue.
When an upstream request arrives, we dynamically split the task based on batch size in descending order, assign it to the corresponding executor in the queue, and push the index back to the queue after the computation is completed. 
Figure \ref{fig:dso} illustrates the schematic diagram of the DSO module.

\section{Experiments}

\subsection{Experimental Setups}
\begin{figure}[htbp]
  \centering
  \includegraphics[scale=0.7]{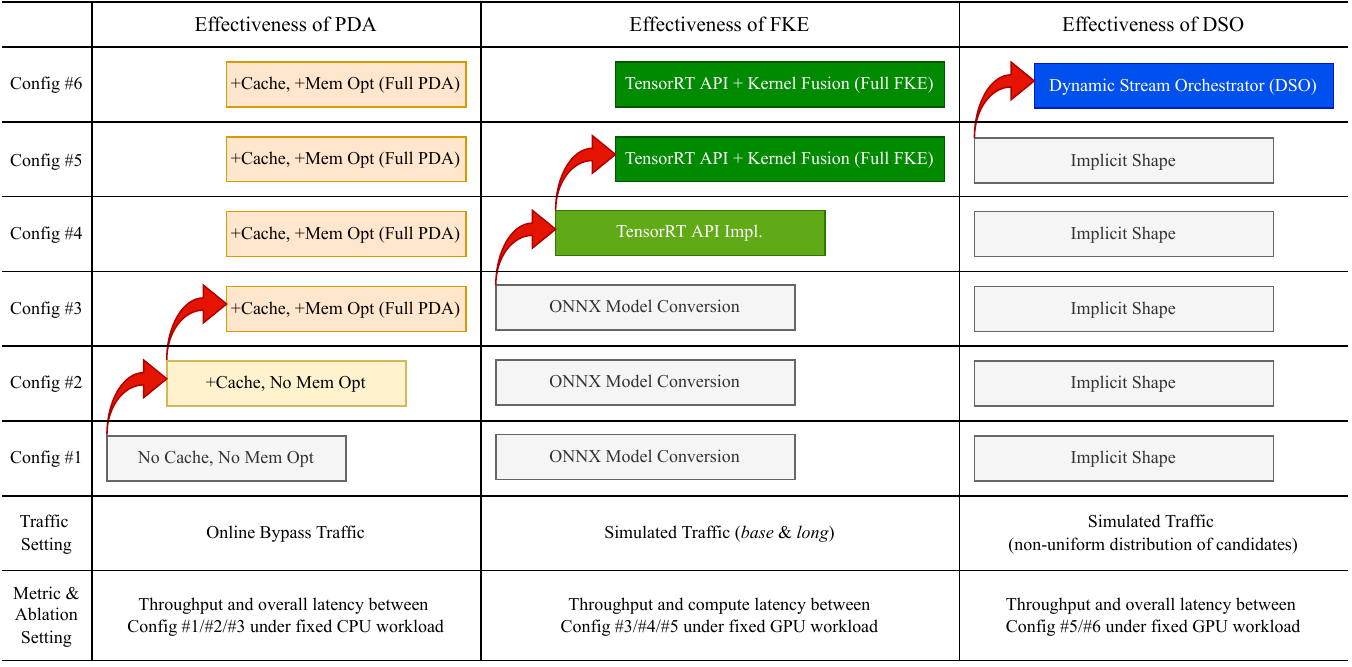}
\caption{Experiment configurations for ablation study.}
\label{fig:exp}
\end{figure}

In this section, we conduct a series of performance tests to demonstrate the effectiveness of each module in our serving system. 
{The experiment configurations for ablation study are shown in Fig \ref{fig:exp}.}
The optimizations are based on the Climber GR model\cite{xu2025efficient}, but the fundamental ideas of these optimizations are equally applicable to other sequence recommendation models dominated by the Transformer architecture. 
In addition to the real-world online model configuration, we also tested two different scenarios with varying historical sequence lengths and numbers of candidate items to approximate the situations that might be encountered online, which we respectively denoted by \textit{base} and \textit{long} according to their different computational loads. 
The relevant statistics are detailed in the Table \ref{table:model}.
Due to its powerful sequence modeling capabilities, the Climber model requires significantly less feature engineering. 
Compared to the hundreds or even thousands of features required by the DLRM model, the GR model in this test primarily focuses on historical interaction sequences on the user side along with only a dozen pieces of side information related to user and item side. 
The system is deployed on containerized CPU-GPU heterogeneous instances, each compromising a CPU with 16 cores and 64GB of memory, and a single NVIDIA 4090D GPU card with 24GB of HBM.
The latency is measured in milliseconds and is categorized into two calculation scopes: end-to-end request processing latency (overall latency) and pure model computation latency.
The former measures the overall scheduling performance of the system, while the latter specifically reflects the performance of the computational kernels.
Throughput is calculated as the number of user-item pairs processed per second. 
Resource utilization corresponds to the computational load and memory usage of the CPU and GPU.

% In the subsequent evaluations, we increased the number of concurrent requests to push the system's capacity to its limit and recorded the system performance metrics such as latency, throughput, and resource consumption under this extreme condition.
% Specifically, the limit state is defined as: (1) failure rate less than 0.05\%, (2) request latency P99.5 less than 100ms, (3) GPU queue discard rate less than 0.5\%.

\begin{table}[htb]
 \centering \caption{Model details corresponding to the two test scenarios}
\begin{tabular}{cccccc}
    \hline
    Scenarios & User Sequence Length & \makecell{\#Candidate Items} & \#Blocks & \makecell{\#Layers per Block} & FLOPS \\
    \hline
    \textit{base} & 512 & 128 & 2 & 12 & $3.72\times 10^9$\\
    \textit{long} & 1024 & 512 & 2 & 12 & $1.64\times 10^{10}$\\
    \hline
\end{tabular}
\label{table:model}
\end{table}

\subsection{Evaluation Results}

\subsubsection{Effectiveness of PDA}

In this part, we focus on the ablation study on the effectiveness of PDA module. 
We conducted performance tests using a bypass stream of real online traffic to evaluate the effectiveness of the caching mechanism during feature querying.
To ensure a fair comparison across versions, we kept CPU utilization at the same level, approximately 16\%.
The key metrics assessed include throughput, overall latency, P99 latency, and network utilization with CPU load held at the same level, providing a comprehensive view of how each configuration influences the module's operation.
The results are presented in Table \ref{table:pda}. 
``Mem Opt" is short for NUMA affinity binding and pinned data transfer. 
Throughput is reported in ``k", denoting thousands of user-item pairs processed per second.
% The ``w/o Cache" and ``w/o Mem Opt" configurations serve as baselines, while the Full PDA (integrating both Cache and Mem Opt) is the optimized version.

Compared to the ``--Cache, --Mem Opt" baseline, the ``+Cache, --Mem Opt" achieves remarkably positive improvements: 
{throughput surges by 57.9\% (from 67.4 k to 106.4 k), a substantial leap that greatly enhances processing efficiency; 
overall latency plummets by 18.1 \% (from 22.6 ms to 18.5 ms), a dramatic reduction that makes the system far more responsive; P99 latency decreases by 12.5\% (from 60 ms to 52.5 ms), ensuring more stable performance for the vast majority of users; 
and network utilization drops by 38.2\% (from 46.3 MB/s to 28.6 MB/s), a welcome decrease that eases network pressure. }
These results fully demonstrate that the caching mechanism reduces network overhead and boosts overall data flow throughput.

Against the ``+Cache, --Mem Opt" version, the ``Full PDA" also shows significant and beneficial enhancements: 
{throughput increases by 19.0\% (from 106.4 k to 126.6 k), further boosting the system's processing capacity; 
overall latency reduces by 28.6\% (from 18.5 ms to 13.2 ms), a notable improvement in speed; 
P99 latency falls by 12.4\% (from 52.5 ms to 46 ms), making the system more reliable.
Meanwhile, network utilization rises by 18.9\% (from 28.6 MB/s to 34 MB/s), but this minor increase is a negligible trade-off given the substantial gains in other key metrics.}
These results confirm that the Full PDA, as the optimized version, outperforms both baselines comprehensively in a highly favorable manner, highlighting the excellent synergistic effect of integrating Cache and Mem Opt.

\begin{table}[htb]
 \centering \caption{Ablation study on PDA module. ``Mem Opt" is short for the combination of NUMA affinity binding and pinned data transfer. Throughput is reported in ``k", denoting thousands of user-item pairs processed per second.}
\begin{tabular}{lccccc}
    \hline
    Ablation Study & Throughput & Overall Latency & P99 Overall Latency & Network Utilization \\
    \hline
    --Cache, --Mem Opt          & 67.4 k  & 22.6 ms & 60 ms   & 46.3 MB/s\\
    +Cache, --Mem Opt           & 106.4 k & 18.5 ms & 52.5 ms & 28.6 MB/s\\
    +Cache, +Mem Opt (Full PDA) & 126.6 k & 13.2 ms & 46 ms   & 34 MB/s\\
    \hline
\end{tabular}
\label{table:pda}
\end{table}

\subsubsection{Effectiveness of FKE}
In this part, we focus on the evaluation of performance induced by optimization for model computation under the two scenarios (\textit{base} and \textit{long}) as mentioned above.
We compare three configurations when building inference engine: ONNX model conversion as the baseline, network re-building via TensorRT API as the advanced version, and replace Transformer implementation with fused-kernel plug-ins as the final optimized version.
The quantitative results of our ablation study (Table \ref{table:fke}) demonstrate progressive performance improvements across optimization stages. 

In the \textit{base} condition, the basic implementation (ONNX Model Conversion) achieves a throughput of 180.5 k, with a compute latency of 5.69 ms and a P99 latency of 8.16 ms, performance that is functional but far from optimal. 
The advanced implementation (TensorRT API Impl.) demonstrates significantly better performance, with throughput increasing by 226.2\% to 588.8 k, compute latency decreasing by 73.5\% to 1.51 ms, and P99 latency dropping by 81.1\% to 1.54 ms, which marks a substantial leap in efficiency and responsiveness. 
The final optimized version (TensorRT API Impl. with the kernel fusion plugin) further enhances performance, with throughput rising by 43.3\% to 843.8 k compared to the advanced implementation, compute latency reducing by 17.9\% to 1.24 ms, and P99 latency falling by 16.9\% to 1.28 ms, which solidifies its superiority in handling the \textit{base} workload.

Under the \textit{long} condition, the basic implementation (ONNX Model Conversion) yields a throughput of 220.2 k, accompanied by a compute latency of 18.57 ms and a P99 latency of 18.94 ms, results that are noticeably sluggish for larger input sizes. 
The advanced implementation (TensorRT API Impl.) shows substantial improvements, with throughput surging by 243.6\% to 756.7 k, compute latency cutting by 70.4\% to 5.49 ms, and P99 latency lowering by 70.6\% to 5.56 ms, making it far more viable for extended workloads. 
The final optimized version (TensorRT API Impl. with the kernel fusion plugin) achieves even greater gains, with throughput jumping by 82.6\% to 1381.4 k relative to the advanced implementation, compute latency dropping by 44.8\% to 3.03 ms, and P99 latency falling by 43.9\% to 3.12 ms, an impressive performance boost that makes it exceptionally well-suited for the demanding \textit{long} condition.

\begin{table}[htbp]
  \centering
  \caption{Ablation study on FKE module. ``Kernel Fusion" stands for the combination of mask-aware flash-attention plug-in and fused-FFN plug-in. Throughput is reported in ``k", denoting thousands of user-item pairs processed per second.}
  \begin{tabular}{llccc}
    \toprule
    Scenario & Ablation Study & Throughput & Compute Latency & P99 Compute Latency \\
    \midrule
    \multirow{3}{*}{\makecell{\textit{base}\\(512+128)}} & ONNX Model Conversion & 180.5 k & 5.69ms & 8.16ms \\
    & TensorRT API Impl. & 588.8 k & 1.51ms & 1.54ms \\
    & TensorRT API Impl. + Kernel Fusion & 843.8 k & 1.24ms & 1.28ms \\
    \midrule
    \multirow{3}{*}{\makecell{\textit{long}\\(1024+512)}} & ONNX Model Conversion & 220.2 k & 18.57ms & 18.94ms \\
    & TensorRT API Impl. & 756.7 k & 5.49ms & 5.56ms \\
    & TensorRT API Impl. + Kernel Fusion & 1381.4 k & 3.03ms & 3.12ms \\
    \bottomrule
  \end{tabular}
  \label{table:fke}
\end{table}

Overall, the ``TensorRT API Impl." outperforms the ``ONNX Model Conversion" significantly across both operating conditions, effectively addressing the latter’s shortcomings. The integration of the fused-kernel plug-ins in the final optimized version further enhances performance, with particularly notable improvements observed in the \textit{long} condition, highlighting the value of the fused-kernel plug-ins in handling more complex workloads.
{Another noteworthy point is that, regardless of the implementation, throughput under the \textit{long} condition consistently exceeds that of the \textit{base} condition.
Qualitatively, when the length of user-history sequence is fixed, simply adding more candidates lowers the average compute overhead per item.
Longer user sequences further increase computational intensity.
Because we count throughput as user–item pairs per second, this amortization directly improves the measured throughput.}

\begin{figure}[htp]
    \centering
    % \subfigure[Throughput]{
    %     \includegraphics[width=0.3\textwidth]{fig/throughput.pdf}
    %     \label{fig:throughput}  
    % }
    % \subfigure[SpeedUp]{
    %     \includegraphics[width=0.3\textwidth]{fig/speedup.pdf}
    %     \label{fig:speedup}  
    % }
    \begin{subfigure}{0.45\textwidth}
        \centering
        \includegraphics[width=\textwidth]{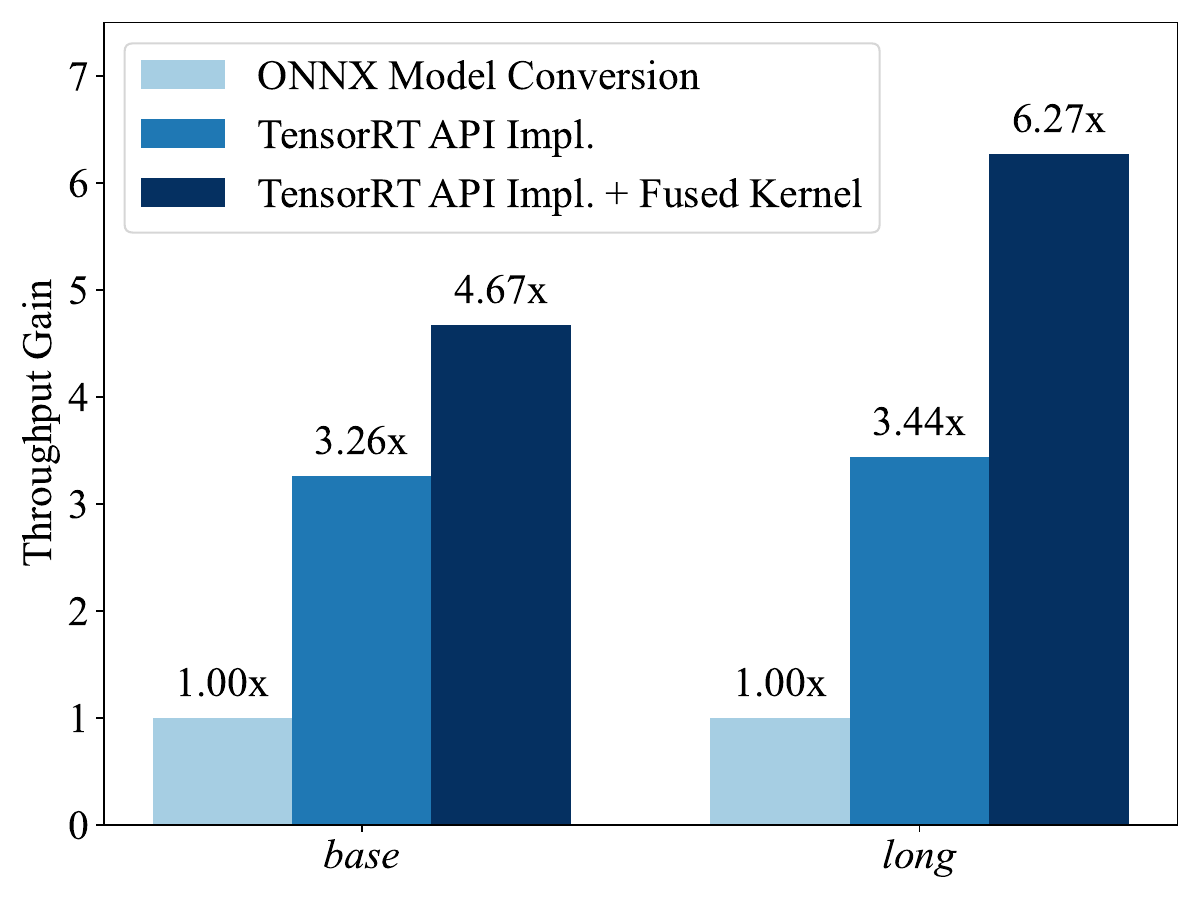}
        \caption{Throughput}
        \label{fig:throughput}
    \end{subfigure}
    \hfill  
    \begin{subfigure}{0.45\textwidth}
        \centering
        \includegraphics[width=\textwidth]{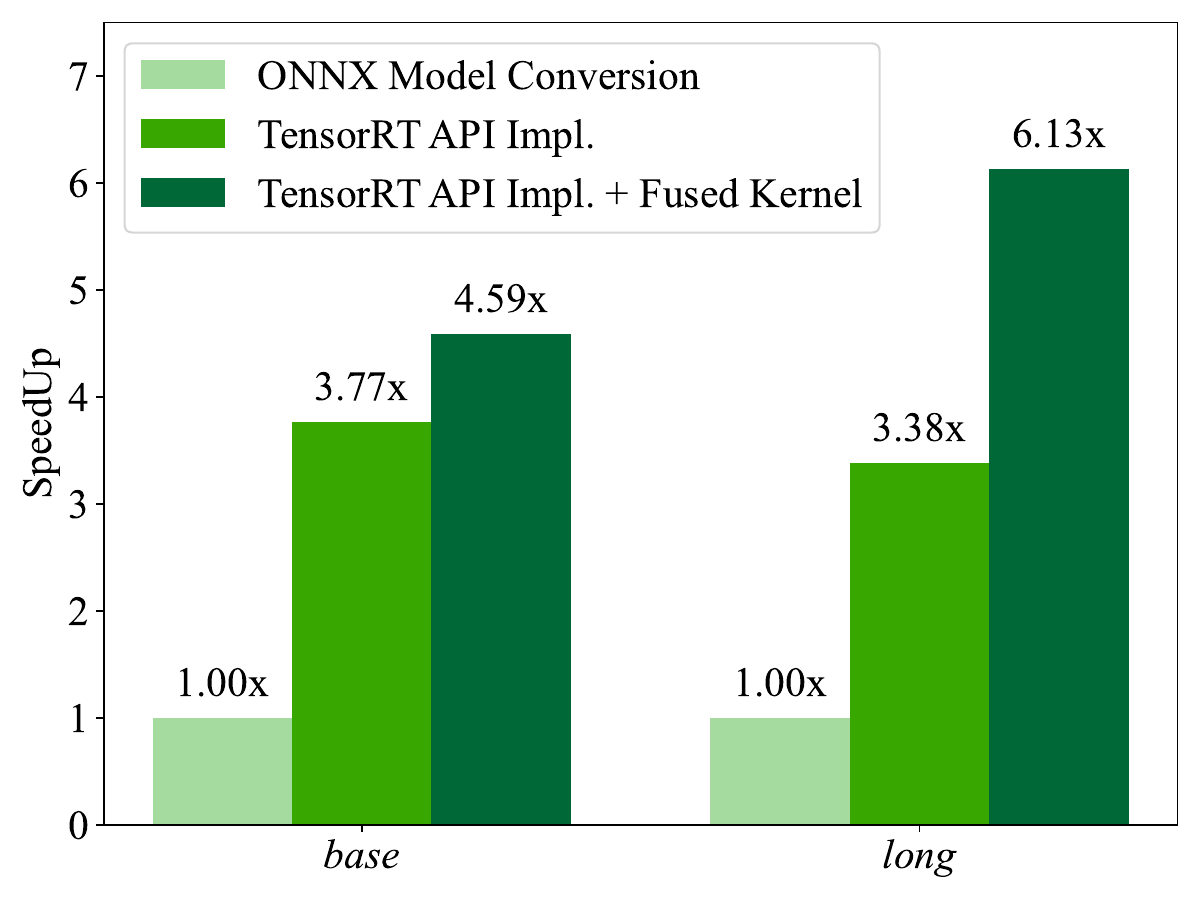}
        \caption{SpeedUp}
        \label{fig:speedup}
    \end{subfigure}
    \caption{Performance comparison among different inference engine building}
    \label{fig:fke_perf}  
\end{figure}

\subsubsection{Effectiveness of DSO}
In this part, we simulate model-computation performance under uneven candidate item distributions from upstream services to evaluate the performance gains of the DSO module in memory optimization and compute scheduling.
We simulated mixed-traffic workloads in which the number of items was uniformly distributed across 128, 256, 512, and 1024. 
{The length of user behavior sequence is fixed as 1024.}

Table \ref{table:dso} presents the results of the ablation study on the DSO module under simulated mixed-traffic workloads, comparing two configurations: Default (Implicit Shape) and DSO (Explicit Shape), with key metrics including throughput, compute latency, and P99 latency.
{The Default (Implicit Shape) configuration serves as a baseline, achieving a throughput of 912 k user-item pairs per second. 
It exhibits an overall latency of 13.6 ms and a P99 latency of 49 ms, performance that, while operational, leaves significant room for improvement in handling mixed-traffic workloads.
In contrast, the DSO (Explicit Shape) configuration delivers remarkable enhancements across all metrics. 
Its throughput surges to 1190.4 k, representing a substantial increase of 30.5\% compared to the baseline. 
This boost in throughput indicates a far more efficient processing capability, making it highly effective for handling the demands of mixed-traffic scenarios. 
Additionally, the overall latency is reduced by 42.6\% to 7.8 ms, and the P99 latency decreases by 28.6\% to 35 ms, reflecting improved responsiveness and stability even under varying workload conditions.}
Overall, the DSO (Explicit Shape) configuration clearly outperforms the Default (Implicit Shape) setup, demonstrating the significant value of adopting explicit shape in optimizing the performance of the DSO module for simulated mixed-traffic workloads.

\begin{table}[htbp]
 \centering \caption{Ablation study on DSO module under simulated mixed-traffic workloads. Throughput is reported in ``k", denoting thousands of user-item pairs processed per second.}
\begin{tabular}{lcccc}
    \hline
    Ablation Study & Throughput & Overall Latency & P99 Latency \\
    \hline
    Default (Implicit Shape) & 912 k   & 13.6 ms & 49 ms\\
    DSO (Explicit Shape)    & 1190.4 k & 7.8 ms  & 35 ms\\
    \hline
\end{tabular}
\label{table:dso}
\end{table}

\subsection{Summary}
\begin{figure}[htp]
    \centering
    \begin{subfigure}{0.48\textwidth}
        \centering
        \includegraphics[width=\textwidth]{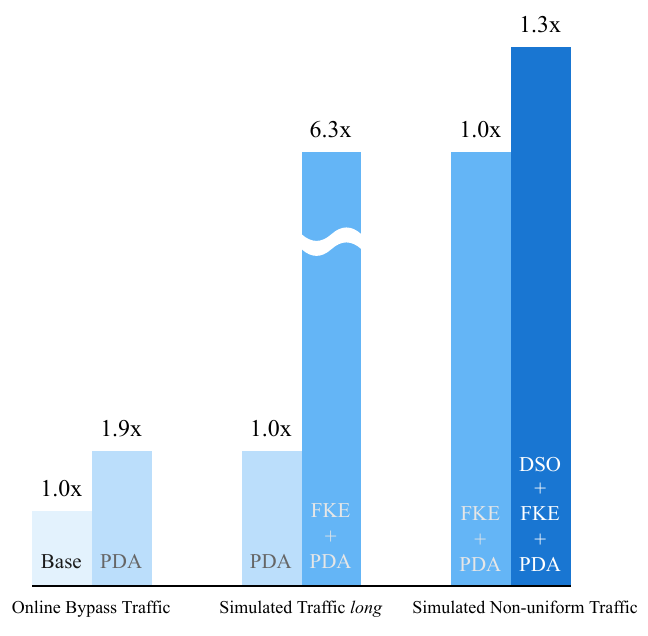}
        \caption{Throughput}
        \label{fig:overall_thp}
    \end{subfigure}
    \hfill  
    \begin{subfigure}{0.48\textwidth}
        \centering
        \includegraphics[width=\textwidth]{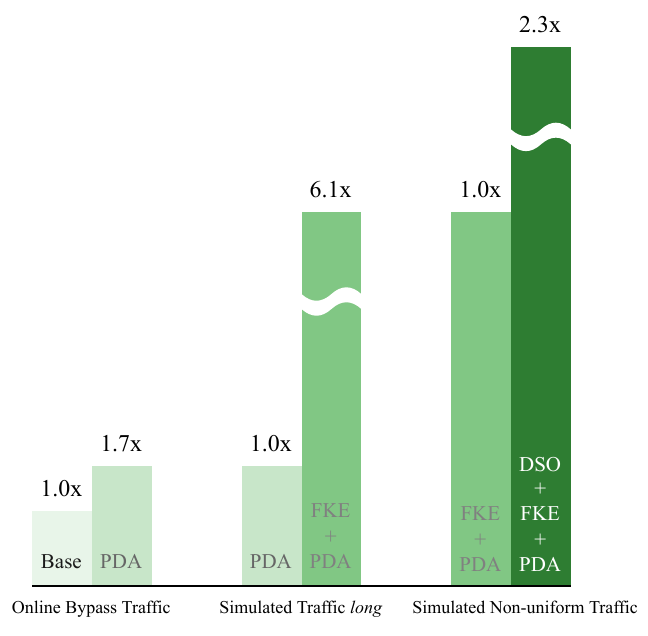}
        \caption{SpeedUp}
        \label{fig:overall_ltc}
    \end{subfigure}
    \caption{Overall performance comparison under different traffic scenarios}
    \label{fig:overall_perf}  
\end{figure}

Overall, through a step-by-step experimental design, we progressively demonstrate the performance of the three proposed modules under their respective traffic scenarios, as illustrated in Fig \ref{fig:overall_perf}.
To evaluate the effectiveness of cached query and memory optimization, we choose online bypass traffic to conduct performance tests. 
Results show that the PDA module can achieve a 1.9x throughput gain and a 1.7x end-to-end request time speed-up comparing to the baseline.
Building on these results, we further validated the critical role of the FKE module in boosting performance under the more demanding \textit{long} workload.
Results show that the FKE module delivers a 6.3× throughput gain and a 6.1× reduction in latency compared with the default engine-building approach of ONNX-to-TensorRT conversion.
What's more, we evaluated the DSO module under simulated traffic with non-uniform upstream candidates, building on the previous findings.
Under this traffic pattern which is closer to real-world conditions, the DSO module delivers 1.3× higher throughput and 2.3× lower latency than the implicit-shape mode that infers batch size at runtime.

\section{Conclusion, Limitation and Future Work}

In this work, we present a comprehensive solution of online serving system tailored \textbf{F}or \textbf{L}arge-scale Gener\textbf{a}tive Reco\textbf{m}mendation with \textbf{E}fficiency (FLAME), leveraging heterogeneous CPU-GPU hardware.
The core innovation lies in the proposed architecture that decouples feature processing from model computation: the CPU handles feature querying, feature extraction and embedding lookup, while the GPU focuses exclusively on model computation, thereby optimizing the division of labor between heterogeneous hardware.
To enhance system throughput, the PDA module is implemented, which encapsulated several memory optimization features. 
{The integration of asynchronous caching mechanism and the utilization of NUMA affinity core binding and pinned data transfer collectively reduce data transmission latency by 1.7x and improve throughput by 1.9x.}
For maximizing GPU computational performance, the Fused-kernel Engine is developed, constructing the main body of the neural network using TensorRT APIs. Specifically, aiming at the prediction paradigm in recommendation tasks, we implemented mask-aware flash attention and perform kernel fusion on the remaining modules of the Transformer to build high-performance plug-ins.
{Compared with building TensorRT inference engine from ONNX model conversion as default, our FKE achieved a speedup ratio of 4.6x-6.1x, throughput gain ratio of 4.7x-6.3x.}
Furthermore, to improve the overall utilization of heterogeneous resources, the DSO module is introduced, which leverages TensorRT's multi-profile optimization and explicit shape mechanisms, employing CUDA streams to achieve dynamic batch routing.
{The DSO module outperforms the default setting (Implicit Shape)  with 1.3x improvement in throughput and 2.3x reduction in latency under non-uniform distribution of upstream candidates.}
In summary, comprehensive evaluations demonstrate that FLAME effectively supports large-scale online deployment of generative recommendation models, achieving remarkable improvements in throughput and latency performance.

{In this work, we didn't adopt the key/value cache mechanism at user level to achieve a two-stage prediction paradigm like M-FALCON \cite{zhai2024actions}.
This choice rests on two considerations.
First, our measurements showed that user-level caching achieved only a modest hit-rate considering the characteristics of music platform recommendation business.
Thus, introducing a caching mechanism for feature queries on the core hot items side offers greater benefits compared to the caching feature on the user side.
Second, with our model settings and workload, the computation saved by the two-stage strategy was roughly canceled out by the additional data transfers needed for the intermediate tensors.
Nevertheless, for long-context inference and larger model sizes, the incorporation of KV-cache may still yield further performance gains.
Accordingly, the design of a distributed KV-cache on GPU along with dynamic eviction/offloading strategies is reserved for future work.
In addition, we will explore more interpretable paradigms for generative recommendation, leveraging the reasoning power of LLMs to unlock greater potential in interactive application scenarios.
}

\bibliographystyle{unsrturl}
\bibliography{sample}

% \begin{thebibliography}{99} %% use BibTeX or add references manually

% \bibitem{krishnan00} E. Krishnan, A. M. Shan, T. Rishi, L. A. Ajith, C. V.
% Radhakrishnan, \textit{On-line Tutorial on \LaTeX{}},
% ``Mathematics'' (Indian \TeX{} Users Group, 2000), \\
% \url{http://www.tug.org/tutorials/tugindia/chap11-scr.pdf}.

% \bibitem{vantrigt97} C. van Trigt, ``Visual system-response functions and estimating reflectance,''
% J. Opt. Soc. Am. A \textbf{14}, 741--755 (1997).

% \bibitem{masters93} T. Masters, \emph{Practical Neural Network Recipes in C++} (Academic, 1993).

% \bibitem{shoop97} B. L. Shoop, A. H. Sayles, and D. M. Litynski, ``New devices for optoelectronics: smart pixels,''
% in \emph{Handbook of Fiber Optic Data Communications},
% C. DeCusatis, D. Clement, E. Maass, and R. Lasky, eds. (Academic, 1997), pp. 705--758.

% \bibitem{kalman76} R. E. Kalman,``Algebraic aspects of the generalized inverse of a rectangular matrix,'' in
% \emph{Proceedings of Advanced Seminar on Generalized Inverse and Applications}, M. Z. Nashed, ed. (Academic, 1976), pp. 111--124.

% \bibitem{craig96} R. Craig and B. Gignac, ``High-power 980-nm pump lasers,''
% in \emph{Optical Fiber Communication Conference}, Vol. 2 of 1996 OSA Technical Digest Series (Optical Society of America, 1996), paper ThG1.

% \bibitem{steup96} D. Steup and J. Weinzierl, ``Resonant THz-meshes,''
% presented at the Fourth International Workshop on THz Electronics, Erlangen-Tennenlohe, Germany, 5--6 Sept. 1996.

% \end{thebibliography}

\end{document}